\documentclass[letterpaper,aps,prb,twocolumn]{revtex4}

\usepackage{amsmath}    
\usepackage{graphicx}   
\usepackage{subfigure}

\begin{document}

\title{Prediction and Measurement of Transient Responses of First Difference Based Chaos Control for 1-dimensional Maps}
\author{Edward H. Hellen}
\email{ehhellen@uncg.edu}
\author{J. Keith Thomas} 
\affiliation{University of North Carolina Greensboro, Department of Physics and Astronomy, Greensboro,
NC 27402}
\date{\today}

\begin{abstract}
Chaotic behavior can be produced from difference equations with unstable fixed points. Difference equations can be used for algorithms to control the chaotic behavior by perturbing a system parameter using feedback based on the first difference of the system value. This results in a system of nonlinear first order difference equations whose stable fixed point is the controlled chaotic behavior. Basing the feedback on the first difference produces distinctly different transient responses than when basing feedback on the error from the fixed point. Analog electronic circuits provide the experimental system for testing the chaos control algorithm. The circuits are low-cost, relatively easy to construct, and therefore provide a useful transition towards more specialized real-world applications. Here we present predictions and experimental results for the transient responses of a first difference based feedback control method applied to a chaotic finite difference 1-dimensional map. The experimental results are in good agreement with predictions, showing a variety of behaviors for the transient response, including erratic appearing non-steady convergence. 


 
\end{abstract}

\maketitle
\section{Introduction}
May showed that the logistic equation, a recursion relation based on a quadratic return map, can generate chaotic behavior.\cite{may}  Proportional feedback methods can control chaos in these systems.\cite{ott,wiener,gauthier,flynn} Typically in these control methods a system parameter is perturbed by an amount proportional to the "error", the difference between the current system value and the unstable fixed point. This control method is simple proportional feedback (SPF). Pyragas introduced an alternative in which the perturbation is proportional to the first difference of the system value (difference between current and previous values), and therefore does not depend on the unstable fixed point.\cite{pyragas} This delayed feedback control (DFC) method successfully controls chaotic behavior in a variety of cases.\cite{socolar98}  Here we apply DFC to a chaotic finite difference return map and derive the mathematical form of the transient response and verify it experimentally with an analog electronic circuit.  

The system whose chaotic behavior is to be stabilized is a 1-dimensional finite difference map for system value $x$ with system parameter $a_0$  
\begin{equation}
\label{eq:returnmap}
x_{n+1}=f\left(x_n,a_0\right).
\end{equation}
Stabilization is accomplished using DFC by perturbing parameter $a_0$ by an amount proportional to the first difference, $d\left(x_n\right)=\left(x_n-x_{n-1}\right)$.  Thus when stabilization is turned on we have a system of two first order nonlinear finite difference equations
\begin{subequations}
\label{eq:system}
\begin{align}
x_{n+1}&=f\left(x_n,a_n\right)\\
a_{n+1}&=a_0+K\left(f\left(x_n,a_n\right)-x_n\right).
\end{align}
\end{subequations}
$K$ is the gain for the feedback that perturbs system parameter $a_n$ from $a_0$. The first difference can be written in terms of the error term
\begin{equation}
\label{eq:firstdiff}
d(x_n)=\Delta x_n-\Delta x_{n-1}
\end{equation}
where the error is $\Delta x=x-x^*$ and the fixed point is $x^*$. Thus DFC has the desired property of the perturbation vanishing when the system value attains the fixed value. Stabilization is attempted only when the magnitude of $d\left(x_n\right)$ is within a specified window for control.


We perform a linear stability analysis to determine the conditions for stabilization of the chaotic behavior and to calculate the form of the transient response. The control algorithm is implemented in an analog electronic circuit.  Measurements from the circuit compare well with predictions. A variety of transient responses are found depending on the gain of the feedback.  Some show a steady convergence of the system value to the fixed value, while other cases appear to converge somewhat erratically. The behaviors are distinctly different from the transient responses obtained when using SPF control based on the error $\Delta x$. The electronic circuit provides a real-world system for testing implementation of the control algorithm. 

    
\section{Predictions}
Here we do linear stability analysis in order to determine the conditions under which chaotic behavior can be stabilized and we find the form of the transient response when stabilization is turned on. We use the first difference $d(x_n)$ as our variable of interest since the perturbation of system parameter $a_n$ is $\Delta a_n=Kd(x_n)$. Using $d(x_n)$ can be useful in real-world applications since the system values are easily obtainable, whereas the fixed value may be unknown or may drift. 

The fixed point of Eq.\ \eqref{eq:system} satisfies $x^*=f(x^*,a_0)$ and $a^*=a_0$. The Jacobian for linearization of Eq.\ \eqref{eq:system} about the fixed point is
\begin{equation}
\label{eq:jacobian}
J=\left( \begin{array}{cc}
f_x & f_a \\
K\left(f_x-1\right)&Kf_a
\end{array} \right) 
\end{equation}
where $f_x$ and $f_a$ are the partial derivatives evaluated at the fixed point.  We point out that knowledge of the fixed point is required to evaluate the partial derivatives.  However during the attempt to control chaotic behavior it is not necessary to monitor and update a potentially drifting fixed point, as long as the partial derivatives do not change too much.  In contrast, a feedback method based on the difference from the system value to the fixed point requires monitoring and updating if the system's fixed point can drift. 

The trace and determinant of the Jacobian are $(f_x+Kf_a)$ and $Kf_a$, respectively. These give the condition for a stable fixed point, 
\begin{equation}
\label{stability}
|f_x+Kf_a|<1+Kf_a<2.
\end{equation}
The right-side inequality requires $Kf_a<1$. We are interested in situations when the fixed point is unstable, so $f_x<-1$. Therefore $Kf_a>0$, and the left-side condition requires $(-f_x-Kf_a)<(1+Kf_a)$. Thus the condition on $Kf_a$ for a stable fixed point is  
\begin{equation}
\label{eq:kfa}
\frac{-f_x-1}{2}<Kf_a<1.
\end{equation}
This gives the result that DFC can not control chaos if the slope of the return map at the fixed point is steeper than $-3$.

Now we find the form of the transient response of the first difference. It follows from Eq.\ \eqref{eq:firstdiff} that Eq.\ \eqref{eq:jacobian} is the Jacobian for $d(x_n)$. Let $y_n=d(x_n)$ and look for solutions $y_n=\lambda^n$. 
Characteristic multipliers are 
\begin{equation}
\label{eq:realroots}
\lambda_\pm=\frac{f_x+Kf_a}{2}\pm\sqrt{\left(\frac{f_x+Kf_a}{2}\right)^2-Kf_a}
\end{equation} 
giving solution 
\begin{equation}
\label{eq:smooth}
y_n=c_1\lambda^n_++c_2\lambda^n_-.
\end{equation}
The coefficients $c_1$ and $c_2$ are determined by two consecutive observed values $y_0=c_1+c_2$ and $y_1=c_1\lambda_++c_2\lambda_-$.  Solving for the coefficients gives
\begin{subequations}
\label{eq:coefs}
\begin{align}
c_1 &=\frac{-\lambda_-y_0+y_1}{\lambda_+-\lambda_-}\\
c_2 &=\frac{\lambda_+y_0-y_1}{\lambda_+-\lambda_-}
\end{align}
\end{subequations}

Setting the discriminant in Eq.\ \eqref{eq:realroots} to zero gives the value of $Kf_a$ for transition from geometric to sinusoidal behavior,
\begin{equation}
\label{eq:transition}
Kf_a=2-f_x-2\sqrt{1-f_x}.
\end{equation}
For values of $Kf_a$ larger than in Eq.\ \eqref{eq:transition} and less than one the characteristic multipliers are complex conjugates and the convergence is sinusoidal.  
Using the Euler representation, $\lambda_{\pm}=re^{\pm i\phi}$, and taking the form of Eq.\ \eqref{eq:smooth} the solution is
\begin{equation}
y_n=c\left(re^{i\phi}\right)^n+c^*\left(re^{-i\phi}\right)^n
\end{equation}
where
\begin{equation}
\label{eq:cmplxcoef}
c=\frac{y_0}{2}+i\left(\frac{y_0Re(\lambda_+)-y_1}{2Im(\lambda_+)}\right)=c_0e^{i\theta}.
\end{equation}
Thus the solution for the first difference is 
\begin{equation}
\label{eq:nonsmooth}
y_n=c_0e^{i\theta}\left(re^{i\phi}\right)^n+c_0e^{-i\theta}\left(re^{-i\phi}\right)^n=2c_0r^n\cos\left(n\phi+\theta\right)
\end{equation}
where 
\begin{equation}
\label{eq:r}
r=\sqrt{Kf_a},
\end{equation}
\begin{equation}
\phi=\arctan\left(\frac{\sqrt{4Kf_a-\left(f_x+Kf_a\right)^2}}{f_x+Kf_a}\right),
\end{equation}
and
\begin{equation}
\theta=\arctan\left(\frac{y_0Re(\lambda_+)-y_1}{y_0Im(\lambda_+)}\right).
\end{equation}
The cosine term can cause erratic appearing convergence distinctly different from the steady geometric convergence for the real multipliers in Eq.\ \eqref{eq:smooth}. 

The result is that we predict convergence (control) for feedback gain $K$ between $\left(1+f_x\right)/\left(-2f_a\right)$ and $1/f_a$, with Eq.\ \eqref{eq:transition} giving the transition from geometric convergence [Eq.\ \eqref{eq:smooth}] to sinusoidal convergerence [Eq.\ \eqref{eq:nonsmooth}]. 

As an example we consider the 1-dimensional H\'{e}non map:
\begin{equation}
\label{eq:Henon}
f\left(x,a\right)=1-ax^2
\end{equation}
with fixed point
\begin{equation}
\label{eq:Henonxfix}
x^*=\frac{-1+\sqrt{1+4a_0}}{2a_0}
\end{equation}
and partial derivatives evaluated at $\left(x^*,a_0\right)$
\begin{equation}
\label{eq:Henonfx}
f_x=-2a_0x^*=1-\sqrt{1+4a_0}
\end{equation}
\begin{equation}
\label{eq:Henonfa}
f_a=-\left(x^*\right)^2.
\end{equation}
For values of $a_0$ between 1.4 and 2 Eq.\ \eqref{eq:Henon} displays a variety of unstable behavior including high period oscillations and chaos.\cite{hellen}  For $a_0 = 1.9$ we find $f_x=-1.93$ and $f_a=-0.259$, so convergence of $d(x_n)$ is predicted for values of $K$ between -1.8 and -3.87 with the transition from geometric to sinusoidal convergence at -1.96.

\begin{figure*}[t]
\begin{center}
\includegraphics[width=0.9\textwidth]{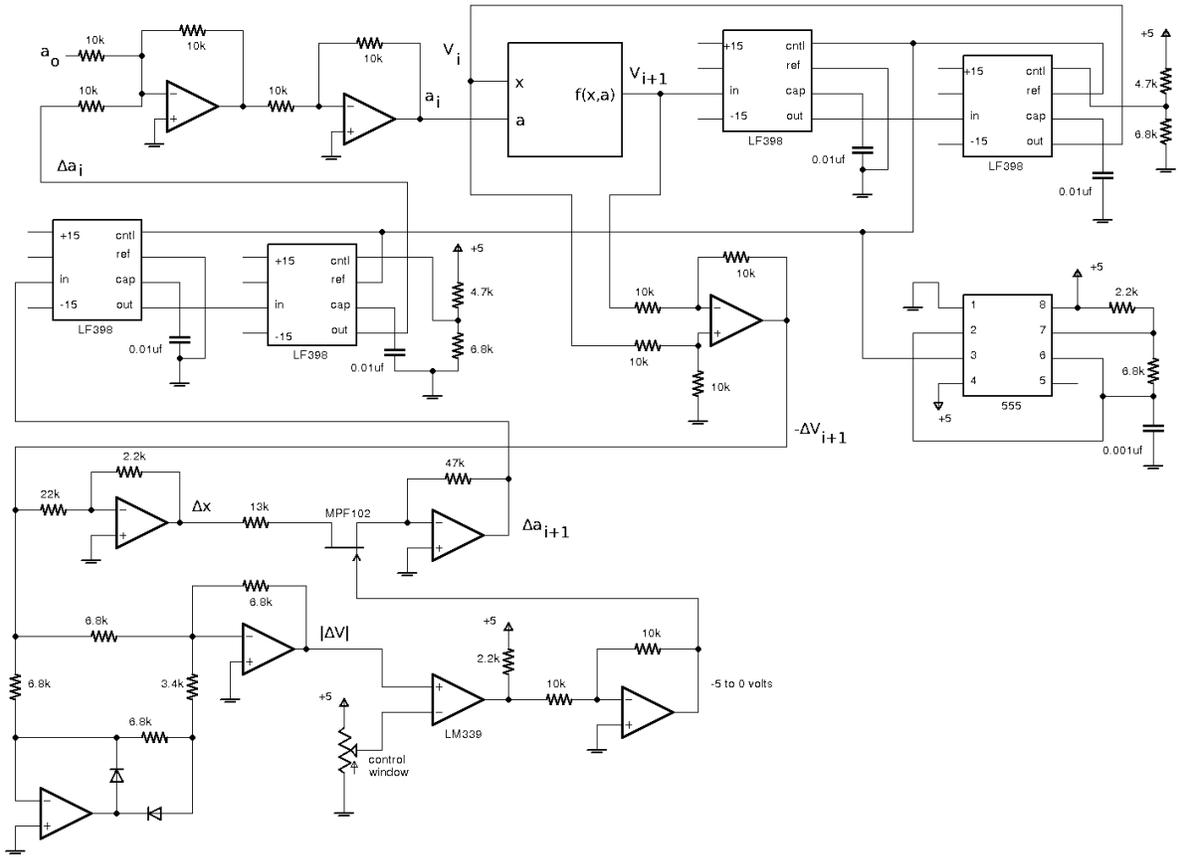}
\caption{\label{fig:circuit}The circuit for controlling chaotic behavior of the return map $x_{n+1}=f(x_n,a_n)$.  Op amps are LF412.}
\end{center}
\end{figure*}  

\section{circuit and measurements}
\begin{figure}[b]
\begin{center}
\includegraphics[angle=0,width=0.45\textwidth]{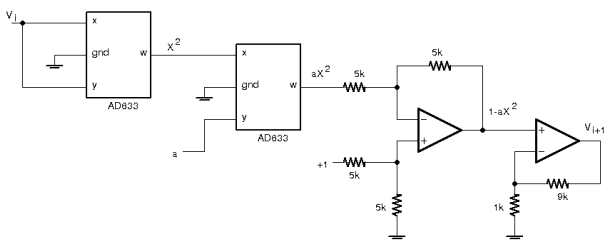}
\caption{\label{function_block}Henon function block circuit for $f\left(x,a\right)$.  Relation between voltage and system value is $x=V/10$. The $\times 10$ noninverting amplifier at the output accounts for both $a$ and the $+1$ not being multiplied by 10, the scaling factor of the AD633 multiplier.}
\end{center}
\end{figure}

Figure \ref{fig:circuit} shows the circuitry used to apply the control algorithm to a function block circuit $f\left(x,a\right)$  that performs analog computation of a chaotic return map.  The voltage $V$ corresponds to the system value $x$, where the scaling factor of the AD633 multiplier integrated circuit must be used so that $x_n=V_n/(10$ volts). Here we use the function block circuit shown in Fig.\ \ref{function_block} that calculates the 1-dimensional H\'{e}non map Eq.\ \eqref{eq:Henon}.\cite{hellen} We have also used function block circuits that produce the Logistic map and the tent map. 

At the upper left in Fig.\ \ref{fig:circuit} the unperturbed parameter value $a_0$ is added to the perturbation $\Delta a_n$ to create system parameter $a_n$.  This is input, along with system value voltage $V_n$, to the $f\left(x,a\right)$ circuit block which produces the next system value voltage $V_{n+1}$.  The subtraction op amp creates the first difference $\Delta V_{n+1}=V_{n+1}-V_n$ that is used to create the perturbation for the next iteration, $\Delta a_{n+1}$.  $\Delta V_{n+1}$ is passed to an absolute value/comparator stage and to a gain stage which produces $\Delta a_{n+1}$.  The output of the comparison stage (LM339) controls the gate of the FET in the gain stage so that if $|\Delta V|$ is larger than the control window then the gate goes to $-5$ volts turning off the FET and thereby setting feedback gain $K=0$.   A nonzero value for $K$ is determined by the inverting op amp adjacent to the FET.  For the values shown, $47k\Omega $ and $13k\Omega$, $K=-3.6$. The sign of $K$ is easily switched by changing the order of inputs $V_n$ and $V_{n+1}$ to the subtraction amplifier.  Prior to the FET $|\Delta V|$ is divided by 10, the scaling factor of the AD633 multiplier used in the $f\left(x,a\right)$ circuit block, to convert from voltage $|\Delta V|$ to $|\Delta x|$. The sample/holds (LF398) act as shift-registers for the iteration of $n$ under the control of the 555 timer circuit. With the $68k\Omega$ and $0.001\mu f$ shown in the schematic the period is about $100\mu s$. 

\begin{figure}[h]
\begin{center}
\includegraphics[angle=0, width=0.5\textwidth]{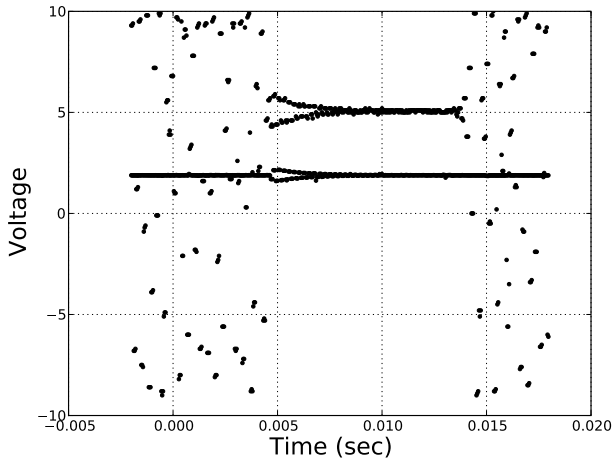}
\caption{\label{fig:smooth_control}Data for control of chaotic H\'{e}non system with feedback gain $K=-1.96$, a value predicted to result in steady convergence of the system value. Control was gated on at $t=0$ and off at $t=0.012$. Also shown is the paramater value $a=a_0+\Delta a$ with $a_0=1.9$.}
\end{center}
\end{figure}
\begin{figure}[h]
\begin{center}
\includegraphics[angle=0, width=0.5\textwidth]{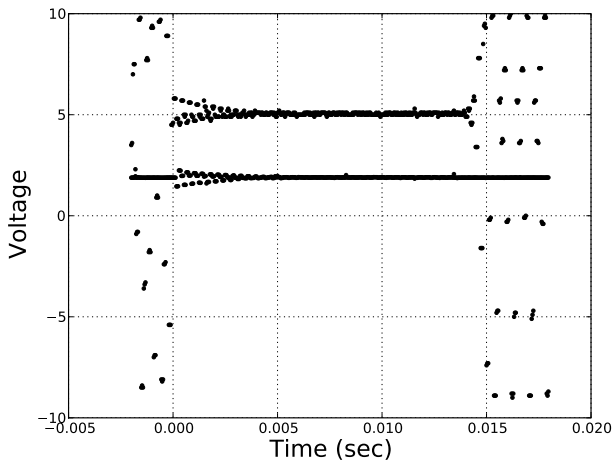}
\caption{\label{fig:nonsmooth_control}Data for control of chaotic H\'{e}non system with feedback gain $K=-3.53$, a value predicted to result in non-steady convergence of the system value. Control was gated on at $t=0$ and off at $t=0.012$. Also shown is the paramater value $a=a_0+\Delta a$ with $a_0=1.9$.}
\end{center}
\end{figure}

Data were collected with a Tektronix TDS 3000 oscilloscope.  The control circuit was periodically gated on and off (circuitry not shown) so that it was possible to trigger from the gating signal in order to capture the entire control of chaotic behavior. Figures \ref{fig:smooth_control} and \ref{fig:nonsmooth_control} show data and the effect of the gating. Also apparent is the control window.  Control was gated on at $t=0$ in both cases, but in Fig.\ \ref{fig:smooth_control} $|\Delta V|$ was not within the control window until about $t=4.5$ ms. 


\section{results and discussion}
Figures \ref{fig:smooth_control} and \ref{fig:nonsmooth_control} show the effectiveness of the control circuit in Fig.\ \ref{fig:circuit} when applied to the H\'{e}non circuit with parameter $a_0 = 1.9$, a value that gives chaotic behavior. The figures show the measured voltages for the system values $V_n=10x_n$ and parameter values $a_n$.  Figure \ref{fig:smooth_control} uses feedback gain $K = -1.96$ corresponding to the transition between geometric and sinusoidal convergence [Eq.\ \eqref{eq:transition}], and Fig.\ \ref{fig:nonsmooth_control} uses $K = -3.53$ corresponding to non-steady sinusoidal convergence. 

\begin{figure}[h]
\begin{center}
\includegraphics[angle=-0, width=0.5\textwidth]{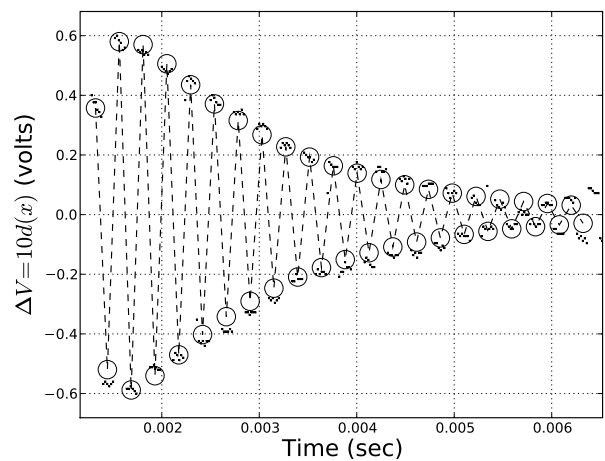}
\caption{\label{fig:smoothdata}Data (dots) and prediction (open circles) showing the first difference of system values $d(x_i)=(x_i - x_{i-1})$ for steady geometric convergence, $K = -1.9$. The connecting dashed lines are for visual aid only.}
\end{center}
\end{figure}

\begin{figure}[h]
\begin{center}
\includegraphics[angle=0, width=0.5\textwidth]{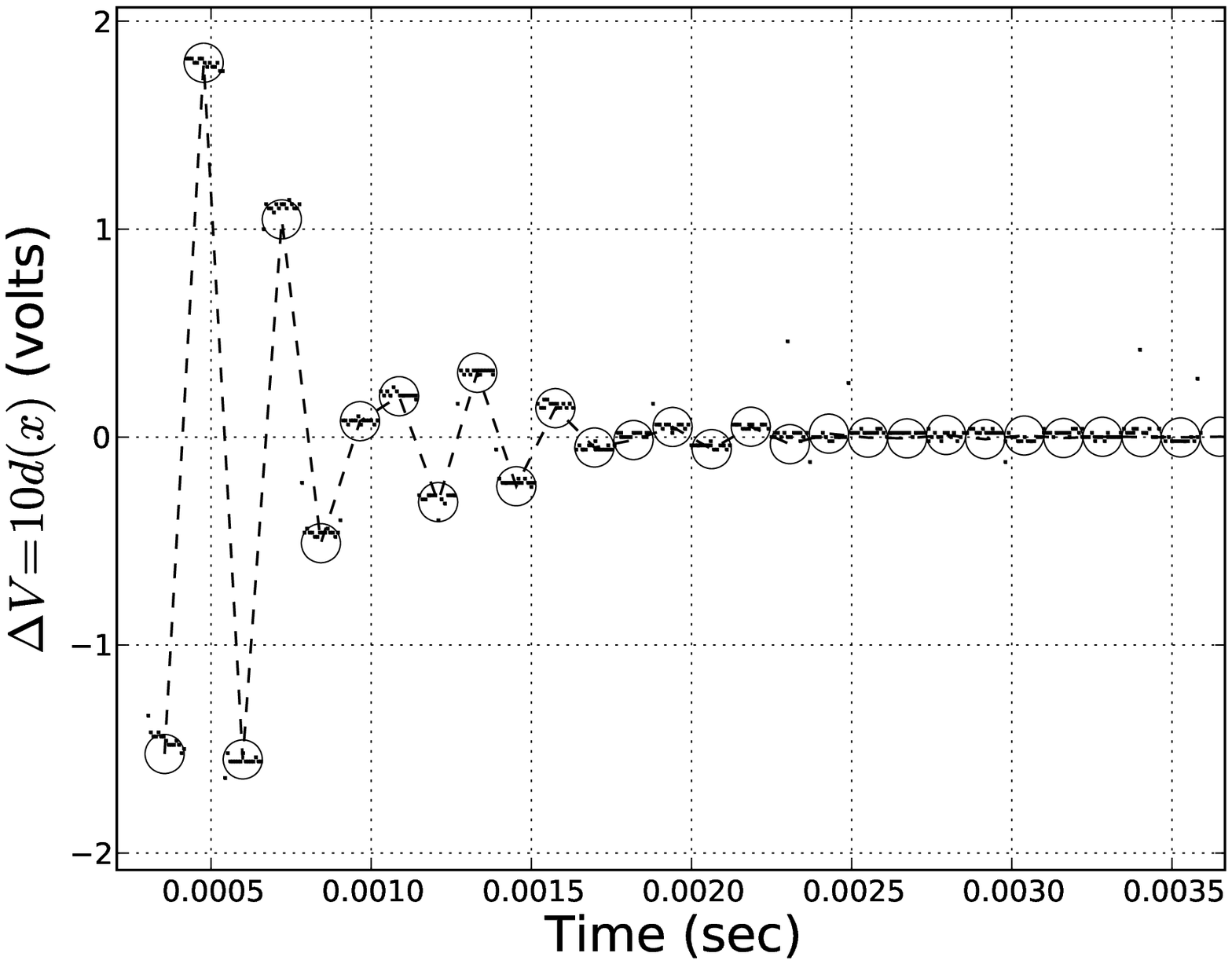}
\caption{\label{fig:transition}Data (dots) and prediction (open circles) showing the first difference of system values $d(x_i)=(x_i - x_{i-1})$ for non-steady sinusoidal convergence, with feedback gain $K = -2.3$. Connecting dashed lines are for visual aid only.}
\end{center}
\end{figure}

\begin{figure}[h]
\begin{center}
\includegraphics[angle=0, width=0.5\textwidth]{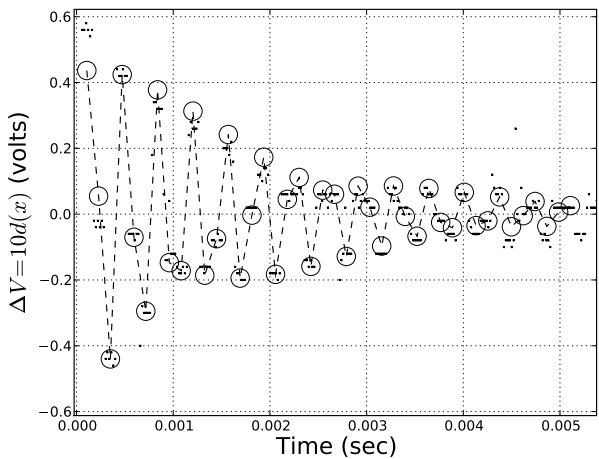}
\caption{\label{fig:nonsmoothdata}Data (dots) and prediction (open circles) showing the first difference of system values $d(x_i)=(x_i - x_{i-1})$ for non-steady sinusoidal convergence, with feedback gain $K = -3.4$. Connecting dashed lines are for visual aid only.}
\end{center}
\end{figure}

The DFC method uses feedback proportional to the first difference of system values, $d(x_n) = x_n-x_{n-1}$. When the system is successfully controlled $d(x_n) \rightarrow 0$. In Section II we showed that the nature of the convergence of $d(x_n)$ depends on the feedback gain $K$. The approach to zero for $d(x_n)$ may be steady [Eq.\ \eqref{eq:smooth}] or may appear somewhat erratic [Eq.\ \eqref{eq:nonsmooth}]. Figures \ref{fig:smoothdata}, \ref{fig:transition}, and \ref{fig:nonsmoothdata} show data and prediction for $y_n$ (actually $\Delta V_n=10y_n$) for three values of $K$ showing the variety of convergence.  Predictions were made by using two successive measured system value differences for $y_0$ and $y_1$ in Eqs.\ \eqref{eq:coefs} and \eqref{eq:cmplxcoef} to determine the coefficients for Eqs.\ \eqref{eq:smooth} or \eqref{eq:nonsmooth}.  The convergence in Fig.\ \ref{fig:smoothdata} is steady, in Fig.\ \ref{fig:transition} it appears somewhat erratic, and in Fig.\ \ref{fig:nonsmoothdata} a pattern is apparent although the convergence is not steady.  In all cases there is good agreement between the prediction and measurement.


When SPF is used to control chaos, system parameter $a$ is perturbed by an amount $\Delta a$ proportional to the error $\Delta x$ from the fixed value. The equations for the system's value and parameter are, 
\begin{subequations}
\label{eq:spf_system}
\begin{align}
x_{n+1}&=f\left(x_n,a_n\right)\\
a_{n+1}&=a_0+K\left(f\left(x_n,a_n\right)-x^*\right).
\end{align}
\end{subequations}
The Jacobian for linear stability analysis is the same as Eq.\ \eqref{eq:jacobian} except for $J_{21}=Kf_x$.  This makes $det(J) = 0$ resulting in stability when $-1-f_x<Kf_a<1-f_x$ and solution $\Delta x_n=A(f_x+Kf_a)^n$. Thus the behavior of the transient response for SPF is quite different from that for DFC.  The SPF transient response has a single characteristic multiplier so it is a steady geometric convergence, either alternating sign or monotonic depending on whether $(f_x+Kf_a)$ is positive or negative. Immediate convergence occurs when $K=-f_x/f_a$ since this causes the characteristic multilplier to be zero.  DFC has either a more complicated geometric convergence governed by two characteristic multipliers, or the sinusoidal convergence that is responsible for erratic appearing non-steady convergence. DFC can not have immediate convergence because according to Eqs.\ \eqref{eq:transition} and \eqref{eq:r} $r$ in Eq.\ \eqref{eq:nonsmooth} can never be zero.  The fastest convergence occurs when $Kf_a$ has its minimal value given by Eq.\ \eqref{eq:transition}. 

For SPF applied to the H\'{e}non map with $a_0=1.9$ the value of $K$ for immediate convergence is $-f_x/f_a=-1.93/0.259=-7.45$. The range of $K$ that give stability is $(-1-f_x)/f_a=-3.59$ to $(1-f_x)/f_a=-11.3$. For DFC, Eq.\ \eqref{eq:kfa} predicts that the range of gain giving stability is $-1.8$ to $-3.87$. This is reasonable since the error $\left(x_n-x^*\right)$ can be expected to typically be about half of the first difference $\left(x_n-x_{n-1}\right)$, so that the SPF method needs a gain about twice that of the DFC gain in order to get a similar perturbation $\Delta a$.  

We have shown that there are a variety of transient responses when applying delayed feedback control to a chaotic system governed by a finite difference 1-dimensional map. In addition, the behaviors of the convergence are distinctly different from the simpler geometric convergence obtained with simple proportional feedback.  Knowledge of this variety is important in situations in which the system's response to application of a control algorithm is being closely monitored since erratic and non-steady convergence could be mistaken for a faulty control mechanism. The important advantage DFC has over SPF is that DFC uses an error signal that does not depend on the unstable fixed point.  This simplifies experimental implementation and also means that DFC will continue to stabilize chaos even if the unstable fixed point drifts by small amounts due to noise or changing system parameters.    
 
\begin{acknowledgments}
This research was supported by an award from the Research Corporation.  Corey Clift contributed to early work on this project. 
\end{acknowledgments}


\newpage

\end{document}